\documentclass[9pt,twocolumnn,a4paper]{scrartcl}

\pdfoutput=1

\usepackage{hyperref}
\usepackage{url}

\usepackage{geometry}
\usepackage{microtype}
\usepackage{authblk}

\usepackage{graphicx}
\usepackage[abs]{overpic}
\usepackage{caption}
\usepackage{subcaption}

\usepackage{cuted}

\usepackage{amsmath, amsthm, amssymb}
\usepackage{amstext}

\usepackage{xcolor}

\usepackage{upgreek}

\newcommand{\doi}[1]{\href{http://dx.doi.org/#1}{\nolinkurl{#1}}}

\newcommand{\onlinecite}[1]{\cite{#1}}

\newcommand{\eref}[1]{(\ref{#1})}
\newcommand{\fref}[1]{FIG.~\ref{#1}}
\newcommand{\tref}[1]{TAB.~\ref{#1}}
\newcommand{\um}{$\upmu$m}

\begin{document}

\title{Experimental evidence of effective $\langle111\rangle$ atomic exchanges in a B2 intermetallic alloy}

%\author{Markus Stana$^1$, Manuel Ross$^1$ and Bogdan Sepiol$^1$}
%\ead{markus.stana@univie.ac.at}

\author{Markus Stana\protect\footnotemark
%\thanks{markus.stana@univie.ac.at}%
, Manuel Ross and Bogdan Sepiol}

\affil{Universit\"at Wien, Fakult\"at f\"ur Physik, Boltzmanngasse 5, 1090 Wien, Austria}

%date{\today}

\twocolumn[

\begin{@twocolumnfalse}

\maketitle

\begin{abstract}
We report on an atomic-scale X-ray Photon Correlation Spectroscopy study on Fe-rich FeAl in the B2-ordered regime revealing effective $\langle111\rangle$-exchanges of atoms on the Al-sublattice. The diffusion mechanism responsible for these exchanges is the dominating process for chemical diffusion and therefore controls long-range material transport in the system. Comparison with previous atomistic results shows that the dominance of further exchanges can be characteristic at moderate temperatures where the system is highly ordered. %Determining atomic jump vectors at temperatures where long-range order in an intermetallic alloy is very high was possible by applying atomic-scale X-ray Photon Correlation Spectroscopy. %The expermientally determined activation energy is in good agreement with other results for the system  
%The modeling of this jump behavior allows to draw conclusion about the specific jump mechanism, e.g.~ruling out the six-jump cycles as dominating process in this system. % Furthermore information about effective vacancy - atom interaction can be gained.  

\end{abstract}

\end{@twocolumnfalse}
]

%\pacs{66.30.-h, 78.70.Ck, 81.05.Bx} 
%\submitto{\JPCM}
% max 4 verwenden

% 66.30.-h Diffusion in solids
% 66.30.Fq Self-diffusion in metals, semimetals, and alloys
% Chemical interdiffusion, 66.30.Ny
% 66.30.Lw Diffusion of other defects

% scattering: x-ray in condensed matter, 78.70.Ck
% materials: metals and alloys, 81.05.Bx
% coherent spectroscopy of atoms and molecules, 82.53.Kp 
% time resolved spectroscopy >1psec, 78.47.D-

%\keywords{XPCS, chemical diffusion, iron-aluminum alloy, intemetallic alloy, short-range order, Synchrotron radiation}

%\maketitle
%\ioptwocol

\footnotetext{markus.stana@univie.ac.at}

\section{Introduction}
\label{sec:intro}
%The goal of this work is to study the movement of single atoms in a B2 ordered intermetallic structure. 
%While finding the preferred paths of atomic motion on a lattice represents an important fundametal question in the dynamics properties research, it can also be interesting for applied physics. 
It is one of the fundamental questions in the dynamics properties research to find the preferred paths of atomic motion on a lattice. Knowledge of such processes can, however, also be interesting for applied physics. Despite attractive properties like high specific strength, good corrosion and oxidation resistance and low material cost, the application of intermetallic compounds is still limited mainly by their high brittleness at low temperatures \cite{Stoloff1994,Herrmann2003}. Knowing movement of single atoms can help in finding a way to tailor materials where such undesired characteristics would be spared.

It is clear that vacancies are the vehicles of diffusion in elementary metals, their solutions and alloys \cite{Mehrer2007}.
While diffusion via a jump mechanism is well understood in solid solutions in frame of the encounter model \cite{Eisenstadt1963,wolf1980diffusion} and was recently investigated with the atomic-scale X-ray Photon Correlation Spectroscopy (aXPCS)\cite{Stana2013}, diffusion in intermetallic alloys is still subject of debate. The important boundary condition for diffusion in ordered intermetallics is the conservation of long-range order \cite{lidiard1957vacancy}. Atomic exchanges of atoms belonging to different sublattices, e.g.~nearest-neighbor jumps along a $\left\langle\frac{1}{2}\frac{1}{2}\frac{1}{2}\right\rangle$-direction in case of a B2 ordered system are therefore suppressed. Effective jumps which lead to atomic exchanges farther than a nearest-neighbor position can not be described by a random movement of vacancies. 
The most established diffusion mechanisms conserving the order of the B2 phase and eventually leading to farther effective jumps are the Antistructure-Bridge Mechanism \cite{Kao1993}, the Triple-Defect Mechanism and the Huntington-McCombie-Elcock or 6-Jump Cycle \cite{huntington_priv,PhysRev.109.605,Elcock1959,Huntington1961} mechanism. 

Even though it is not possible to directly follow the movement of the vacancy in an experiment, information about effective exchange vectors can shine light on the matter. Such effective jumps can either result from a direct exchange of an atom and a vacancy (e.g. direct $\langle 100\rangle$-jumps in a B2 system) or as changes in configuration before and after a sequence of correlated jumps like in the encounter model or the mechanisms mentioned above.  

Fe--Al B2 ordered systems have been subject to plenty of diffusion studies (see e.g.~Kao et al.~\cite{Kao1995}). For nearly stoichiometric FeAl as well as for a Fe$_{0.54}$Al$_{0.46}$ B2 system it was found that effective jumps are performed along $\langle100\rangle$ and $\langle110\rangle$-directions with a ratio of $\langle 110\rangle/\langle100\rangle$ of 2:1 with an optional admixture of up to 10\% of $\langle111\rangle$-exchanges \cite{Vogl1994,Feldwisch1995}. 
These distribution could not be related directly to any specific diffusion mechanism. They could, however, be reproduced by Monte Carlo simulations \cite{weinkamer1999monte}. One should notice that this was achieved assuming unreasonably high exchange energies between vacancies and atoms. The particular experimental result was that the dominant jump mechanism at elevated temperatures leads to an effective $\langle110\rangle$-jump.    

In this paper we report on the observation of chemical diffusion in Fe$_{0.54}$Al$_{0.46}$ at much lower temperatures where the degree of B2 order is considerably higher than in previous Quasielastic M{\"o}ssbauer Spectroscopy (QMS) studies \cite{Vogl1994,Feldwisch1995}. Such an investigation was possible by using the new method of aXPCS \cite{Leitner2009}. This method is sensitive to chemical diffusion where atoms of different type exchange position in a coupled way depending on the short-range order in the system. 
This allows determining an effective diffusion coefficient which describes mutual influence of two or more alloy components. 
An effective jump is the exchange of two atoms of different type which in this particular case are iron and aluminum atoms. The aim of this experiment was to find out whether temperature and degree of long range order influence the probability distribution of effective jump vectors.% It could be shown that the dominant jump mechanism at moderate temperatures results in $\langle111\rangle$-exchanges. 

\section{Theory}
\label{sec:theory}
For a thorough introduction to the XPCS method see e.g.~\onlinecite{Sutton2008, Leitner2012, Stana2014}. Here we give only a very brief outline of the theory. 

The intermediate scattering function $g^{(1)}\left(\vec{q},\Delta t\right)$ is the Fourier transform of the Van Hove correlation function and therefore contains like the original function, complete information about atomic motion in space and in time.
 
The Siegert relation connects the intensity autocorrelation function, which is measured in an XPCS experiment and the amplitude correlation function, which is equivalent to the intermediate scattering function:
\begin{equation}
 g^{(2)}\left(\vec{q},\Delta t\right) = \frac{ \left\langle I(\vec{q},t)I(\vec{q},t+\Delta t) \right\rangle_t}{\left\langle I(\vec{q},t) \right\rangle_t^2} = 1 + \beta \left[g^{(1)}\left(\vec{q},\Delta t\right)\right]^2 
\end{equation}
Here, $\vec{q}$ is the scattering vector, $\langle...\rangle_t$ is the time-ensemble average and $\beta$ is the contrast or Siegert factor \cite{Shpyrko2014} and accounts for the coherent part of X-ray scattering. 

In case of a discrete motion on a Bravais lattice the intensity autocorrelation function can be fitted as a simple exponential decay
 \begin{equation}
 g^{(2)}\left(\vec{q},\Delta t\right) = 1 + \beta \exp\left[-2\frac{\Delta t}{\tau(\vec{q}\,)}\right] .
\label{eq:g2fit}
\end{equation}
The correlation time $\tau(\vec{q}\,)$ connects the measured data with the theoretical diffusion model in the following way:
\begin{equation}
 \tau(\vec{q}\,) = \tau_0 \frac{I_\text{SRO}(\vec{q})}{\sum_n p_n\sum_{\Delta\vec{r}_{n_j}} \Big[1-\exp\left(\text{i} \vec{q} \cdot \Delta\vec{r}_{n_j}\right) \Big]}
\label{eq:tau}
\end{equation}
Here, $\tau_0$ is the average residence time of an atom, $p_n$ is the probability for an atom to end up at a particular neighboring position in the $n$-th neighboring shell after an exchange and $\Delta\vec{r}_{n_j}$ is the vector that points to the $j$-th atom position in the $n$-th shell. The scaling term of short-range order intensity $I_\text{SRO}(\vec{q})$ accounts for De
Gennes narrowing \cite{DeGennes1959} derived in this form for crystalline solids by Sinha and Ross \cite{sinha1988self} and is necessary as aXPCS is a coherent method \cite{Leitner2011}. 

In ordered alloys with different sublattices $\sigma$, effective jumps between the sublattices can take place. In a B2 structure with two simple cubic sublattices $\sigma^A$ and $\sigma^B$, the following effective jumps can occur: $\sigma^A\rightarrow\sigma^A$, $\sigma^A\rightarrow\sigma^B$, $\sigma^B\rightarrow\sigma^A$ and $\sigma^B\rightarrow\sigma^B$. 
%Each of these jump processes has a correlation time $\tau^{XY}(\vec{q}\,)$ with $X,Y \in \{A,B\}$.
For a B2 system, where the number of thermal defects is much smaller than the number of constitutional defects, it is possible that one sublattice is completely filled with the majority atoms, i.e.~Fe ($A$-atoms). In this particular case the other sublattice $\sigma^B$ is filled with the remaining excess atoms of type $A$ and with $B$ atoms. If so, chemical diffusion takes place only on the simple cubic Bravais lattice $\sigma^B$ according to \eref{eq:tau}. This defect type with anti-structure atoms on one sublattice is often referred to as triple-defect system \cite{leitner-def-2015}. 

\section{Experimental}
\label{sec:exp}

The experiment was carried out at beamline P10 at the Petra III synchrotron source in Hamburg. The energy of the incident X-ray beam was chosen to be 7.05\,keV in order to avoid iron fluorescence. The single crystalline sample was oriented with the [111]-direction parallel to the beam. The parameterization of the $\vec{q}$-vector by the scattering angle $2\Theta$ and an azimuthal angle $\varphi$ is given in the appendix. $\varphi=0$ corresponds to the $[\bar{1}\bar{1}2]$-direction.  
 
\begin{figure}%[b!]
 	\centering
	 \begin{overpic}[width=0.5\textwidth]{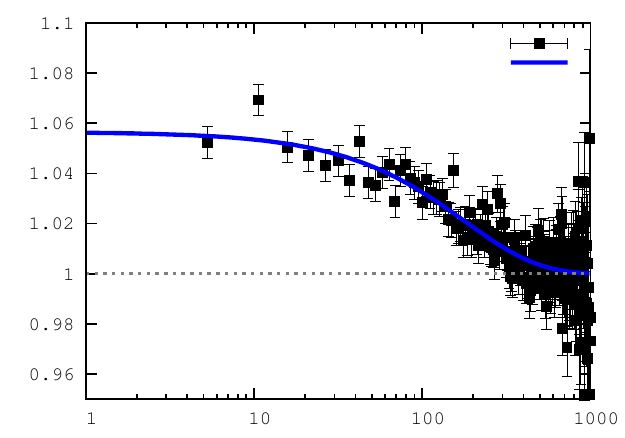}
			\put(126,-4){\small $\Delta t$ (s)}
			\put(-5,74){\rotatebox{90}{\small $g^{(2)}(\vec{q},\Delta t)$}}
			\put(101,153){\tiny $\left\langle I(\vec{q},t)I(\vec{q},t+\Delta t) \right\rangle/\left\langle I(\vec{q},t) \right\rangle^2$}
			\put(126,145){\tiny $1 + \beta \exp\left(-2\Delta t/\tau(\vec{q}\,)\right)$}
 	\end{overpic}	
 	\caption{Intensity autocorrelation function measured at a scattering angle $2\Theta=10^\circ$ and an azimuthal angle \hbox{$\varphi=2^\circ$}. The fit according to \eref{eq:g2fit} yields a correlation time of \hbox{$\tau(\vec{q}\,)=361(22)$\,s} and $\beta = 0.056(3)$. }  
	\label{fig:acf}
\end{figure}
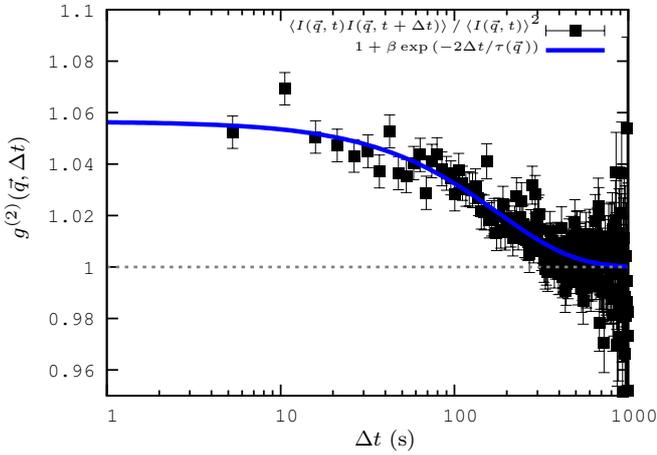

A distance  between sample and detector was 83\,cm. A PIXIS-XB with a CCD array of 1340$\times$1300 pixel of 20$\times$20 \um$^2$ size was used. This camera had a readout time of 2.29\,s/frame and an exposure time of 3\,s was chosen. In order to monitor the stability of the setup and to check the equilibrium state of the sample, a two-time correlation function was calculated for each set of frames taken at a particular $\vec{q}$-vector and measurements were repeated in arbitrary order for  different $\vec{q}-$vectors. A typical intensity autocorrelation function is shown in \fref{fig:acf}.  

The single crystalline sample  was purchased from Lamprecht, Germany (now MaTecK) and was the same sample as used in a QMS study by Feldwisch et al. \cite{Feldwisch1995}. The composition was examined by EDX 
and was found to be to $c_\text{Fe}$ = 0.539(7) and $c_\text{Al}$ = 0.460(7). The lattice constant in Fe$_{0.54}$Al$_{0.46}$ is 2.929\,\AA~at room temperature \cite{Taylor1958}. 

\section{Data evaluation}
\label{sec:dataev}

In order to evaluate the measured dataset with respect to \eref{eq:tau}, the short-range order intensity of the excessive iron atoms on the Al-sublattice has to be known. Unfortunately, information about short-range order in ordered alloys is very scarce in the literature and no information about the system at hand could be found. Attempts to experimentally determine the very weak diffuse scattering intensity were inconclusive. 
Therefore a simulation approach was chosen to determine the Warren-Cowley short-range order parameters \cite{Cowley1960} for the Al-sublattice, which in turn can be used to calculate the corresponding short-range order intensity:
\begin{equation}
 I_\text{SRO}(\vec{q}) = \sum_{n,j} \alpha^{\text{Al}}_n \cos\left(\vec{q} \cdot \Delta\vec{r}_{n_j} \right) 
\label{eq:isro}
\end{equation}

The short-range order parameters are defined as $\alpha^{\text{Al}}_n = 1-w^{\text{Al}}_n/c^{\text{Al}}$, where $w^{\text{Al}}_n$ is, starting from a Fe atom on the Al-sublattice, the probability to find an Al atom in the $n$th neighboring shell within the Al-sublattice and $c^{\text{Al}}$ is the overall concentration of Al atoms on the Al-sublattice.

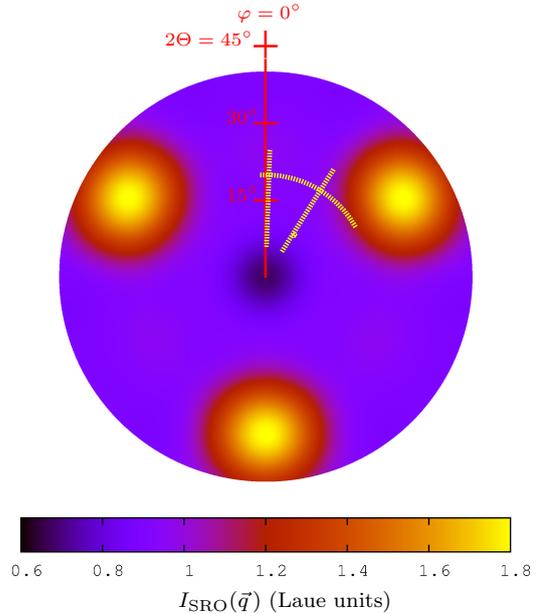
\begin{figure}[!b]
 	\centering
	\begin{overpic}[width=0.4\textwidth]{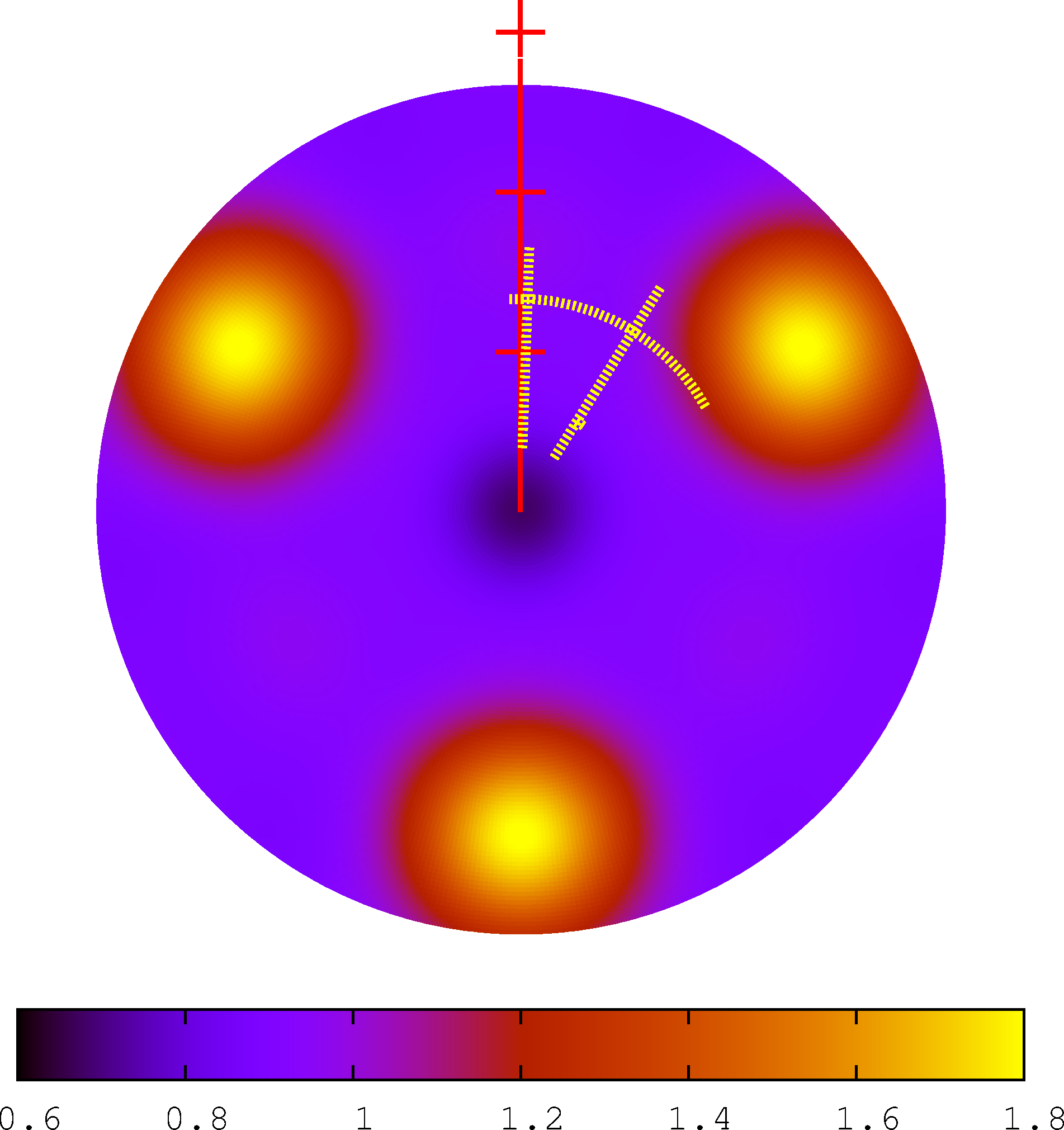}
		\put(62,-12){\small $I_\text{SRO}(\vec{q}\,)$ (Laue units)}
		\put(80,141){\footnotesize {\color{red}{15$^\circ$}}}
		\put(80,170.4){\footnotesize {\color{red}{30$^\circ$}}}
		\put(57,200){\footnotesize {\color{red}{$2\Theta = 45^\circ$}}}
		\put(84,210){\footnotesize {\color{red}{$\varphi = 0^\circ$}}}
	\end{overpic}
	\vspace*{0.4cm}
	\caption{A map of short-range order intensity calculated according to \eref{eq:isro} for a sample oriented in [111]-direction from Warren-Cowley parameters as given in \tref{tab:alphas}. The yellow lines represent the regions of reciprocal space where aXPCS measurements were carried out (cf.~\fref{fig:measurements}). The azimuthal angle is chosen such that $\varphi=0$ corresponds to $[\bar{1}\bar{1}2]$-direction.}  
	\label{fig:isro}
\end{figure}

Simulations were carried out using an off-lattice Monte-Carlo approach \cite{Berthier2007}. The potentials used were taken from an embedded atom method study by Ouyang et al. \cite{Ouyang2012}. This potential reproduces successfully numerous properties of the Fe--Al system, like phonon dispersion spectra and elastic constants. For a temperature of $0.42\times$T$_m$ a vacancy concentration of 0.2\% was used due to experiment \cite{Kogachi1997,Hehenkamp2001}. For a detailed discussion of the simulation procedure see \onlinecite{Stana2016-sim}. With these potentials a lattice constant of $a=$2.98\,\AA~was simulated, which is in reasonable agreement with experimental data \cite{Taylor1958}. The Warren-Cowley parameters found with this approach are given in \tref{tab:alphas}. %One should notice rather weak degree of ordering what explains impracticality of experimental determination of these parameters.

\begin{table}[t]
{\footnotesize
\begin{center}
\begin{tabular}{|l|c|c|c|c|c|c|c|}
\hline
$n$ 						& 1		& 2 		& 3 		& 4 		\\ \hline
$l m n$ 					& 100 		& 110 		& 111 		& 200 		\\ \hline
$\alpha^{\text{Al}}_n$\,$\times 10^{-2}$	& -6.15(4) 	& 2.33(6) 	& -1.45(5) 	& 0.16(6) 	\\ \hline\hline 
$n$ 						& 5		& 6 		& 7		& 8 		\\ \hline
$l m n$ 					& 210 		& 211 		& 220 		& 221 		\\ \hline
$\alpha^{\text{Al}}_n$\,$\times 10^{-2}$	& -0.44(3) 	& -0.08(4) 	& -0.36(5) 	& 0.00(4) 	\\ \hline 			
			
\end{tabular}
\end{center}
\caption{Warren-Cowley parameters calculated using an off-lattice Monte Carlo approach \cite{Stana2016-sim} for neighbor \hbox{shells ($n$)} within the cubic Al-sublattice in B2 ordered Fe$_{0.54}$Al$_{0.46}$. 
Components of vectors to a particular neighbor shell in unit of $a$ are given by $l m n$.}
\label{tab:alphas}
}
\end{table}

\section{Results}
\label{sec:results}

\subsection{Determination of effective jumps}

\begin{figure}[pb]
	\begin{overpic}[width=0.5\textwidth]{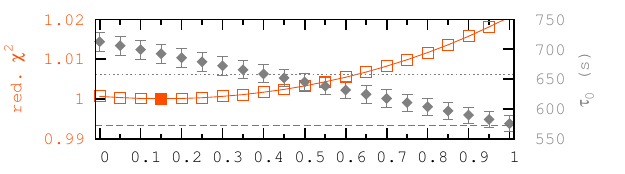}
		\put(108,-5){ $\bar{p}_{100}$ }
		\put(108,73){ $\bar{p}_{110}$ }
		\put(35,63){ \scriptsize 1 }
		\put(195,63){ \scriptsize 0 }
		\put(111,63){ \scriptsize 0.5 }
		\put(221,23){\rotatebox{90}{\colorbox{white}{ {\color{gray} \small $\tau_0$\,(s)}} } }
	\end{overpic}
	\caption{Variance of residuals and average residence time ($\tau_0$) from fitting all measured intensity correlation functions simultaneously with a model consisting of $\langle100\rangle$ and $\langle110\rangle$-exchanges. The best fit was found for 15\% $\langle100\rangle$-exchanges. The broken line represents the variance of residuals and the dotted line $\tau_0$ for a model with 36\% $\langle100\rangle$, 12\% $\langle110\rangle$ and 52\% $\langle111\rangle$-exchanges.}
	\label{fig:model1}
\end{figure}
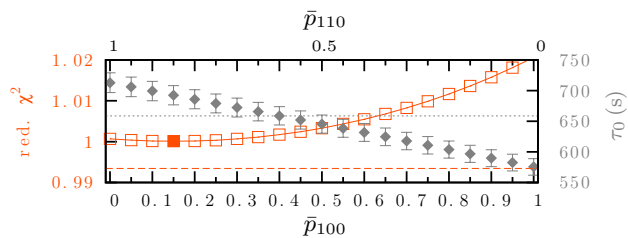

Previous QMS measurements yielded distributions of effective jump vectors for self diffusion in a near stoichiometric B2 system of 35\,\% of $\langle100\rangle$-jumps and 65\,\% of $\langle110\rangle$-jumps or alternatively 26\,\% of $\langle100\rangle$-jumps, 64\,\% of $\langle110\rangle$-jumps and 10\,\% of $\langle111\rangle$-jumps at 1338\,K \cite{Vogl1994}. The same technique was used for studying an off-stoichiometric nominal composition of Fe$_{0.55}$Al$_{0.45}$ \cite{Feldwisch1995}. An effective jump vector distribution of 30\,\% of $\langle100\rangle$-jumps, 50\,\% of $\langle110\rangle$-jumps and 20\,\% of $\langle111\rangle$-jumps was found at a temperature of 1363\,K, which corresponds to a homologous temperature of 0.87$\times$T$_m$.

Here we present an aXPCS measurement carried out at 653\,K, which corresponds to a homologous temperature of only 0.42$\times$T$_m$. Intensity autocorrelation functions were recorded for different pairs of $(2\Theta,\varphi)$. Two models were tested against the correlation times gained according to \eref{eq:g2fit}. The first model implicated a combination of $\langle100\rangle$ and of $\langle110\rangle$-exchanges, that is a combination of nearest-neighbor and next-nearest neighbor effective exchanges within the simple cubic Al-sublattice. For the second model, $\langle111\rangle$-exchanges were additionally included.
All experimentally determined correlation functions of the datasets for points in reciprocal space parameterized by $2\Theta$ and $\varphi$ were fitted simultaneously. The free parameter of this fit according to \eref{eq:tau} were the average residence time $\tau_0$ and a contrast $\beta(2\Theta,\varphi)$ for each pair of adjustable measurement parameters.% resulting in 5493 degrees of freedom.

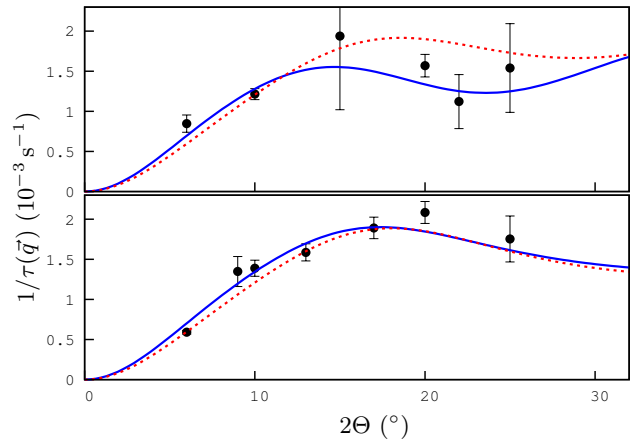
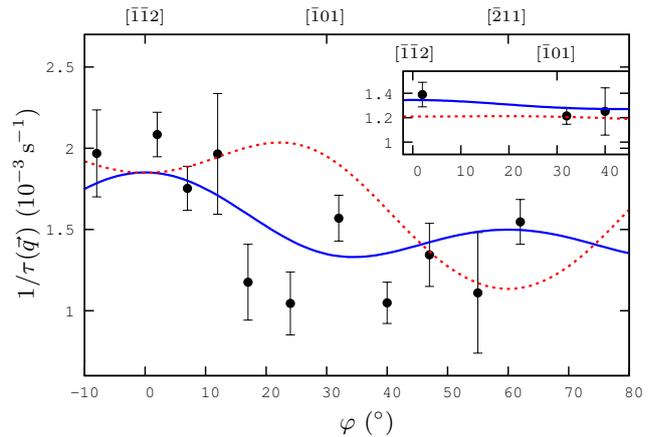
\begin{figure}[t]
	\begin{subfigure}{0.5\textwidth}
 		\begin{overpic}[width=\textwidth]{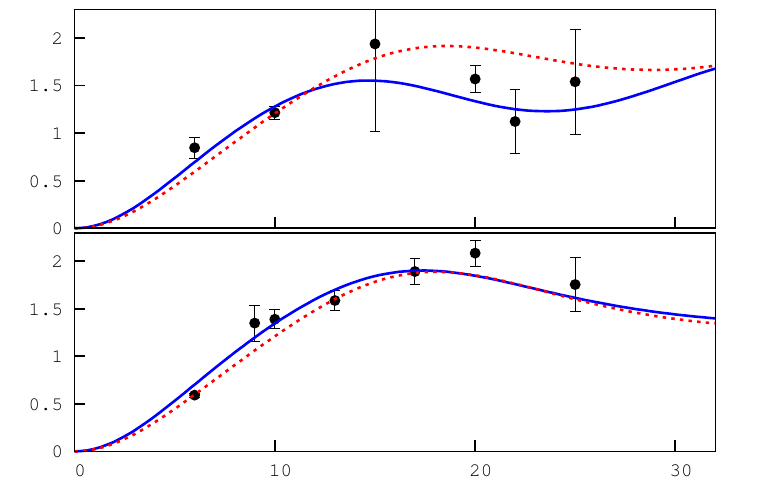}
			\put(116,-8){ $2\Theta$ ($^\circ$) }
			\put(-3,38){\rotatebox{90}{ $1/\tau(\vec{q}\,)$ ($10^{-3}$\,s$^{-1}$)}}
		\end{overpic}
		\vspace*{0.2cm}
 		\caption{Inverse of correlation time $\tau(\vec{q}\,)$ as yielded by a fit of intensity-correlation functions at different scattering angles $2\Theta$ for $\varphi=2^\circ$ (bottom) and $\varphi=32^\circ$ (top).}		
		\label{fig:zwthscan} 
	\end{subfigure}	
	\begin{subfigure}{0.5\textwidth}
		\begin{overpic}[width=\textwidth]{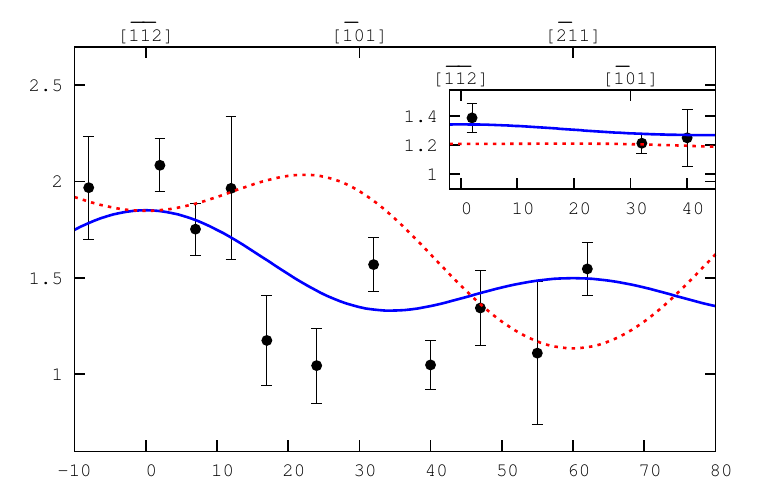}
			\put(116,-8){ $\varphi$ ($^\circ$) }
			\put(-3,38){\rotatebox{90}{ $1/\tau(\vec{q}\,)$ ($10^{-3}$\,s$^{-1}$)}}
			\put(32.5,146){ {\colorbox{white}{\scriptsize $[\bar{1}\bar{1}2]$}} }
			\put(100,146){ {\colorbox{white}{\scriptsize $[\bar{1}01]$}} }
			\put(168,146){ {\colorbox{white}{\scriptsize $[\bar{2}11]$}} }
			\put(131,132){ {\scriptsize \colorbox{white}{ $[\bar{1}\bar{1}2]$}} }
			\put(184,132){ {\scriptsize \colorbox{white}{ $[\bar{1}01]$}} }
		\end{overpic}
		\vspace*{0.2cm}	
		\caption{Inverse of correlation time $\tau(\vec{q}\,)$ yielded by a fit of intensity-correlation functions at different azimuthal angles $\varphi$ for $2\Theta=20^\circ$ and $2\Theta=10^\circ$ (insert). }
		\label{fig:phiscan}
	\end{subfigure}
	\caption{Diffusion models compared with aXPCS measurements. The black symbols with error bars represent the inverse of the correlation times for different positions in reciprocal space. The full, blue line represents a model including 36\% $\langle100\rangle$, 12\% $\langle110\rangle$ and 52\% $\langle111\rangle$-exchanges. The broken, red line represents a model with 15\% $\langle100\rangle$ and 85\% $\langle110\rangle$-exchanges. }
	\label{fig:measurements}
\end{figure}

Results of this procedure are shown in \fref{fig:model1}. For the first model composed of $\bar{p}_{100}$ $\langle100\rangle$-exchanges and of \hbox{$\bar{p}_{110}$ = 1 - $\bar{p}_{100}$} $\langle110\rangle$-exchanges it can clearly be seen, that $\bar{p}_{110} > \bar{p}_{100}$. The best fit was found for 15\% $\langle100\rangle$ and 85\% $\langle110\rangle$-exchanges. It can also be seen that introducing the possibility of $\langle111\rangle$-exchanges yields an improved agreement of model 2 with the experimental data. The best fit for this model was gained combining 36\% $\langle100\rangle$, 12\% $\langle110\rangle$ and 52\% of $\langle111\rangle$-exchanges. 

The fact that $\langle111\rangle$-exchanges are needed to reproduce the measured data can also be seen when visually comparing both models, each with best fitting parameter, with individually fitted correlation times $\tau(\vec{q}\,)$ for each point in reciprocal space where measurements were taken. This is shown in \eref{eq:tau}.

%It shall be noted, that neglecting short-range order in the models ($I_\text{SRO}(\vec{q}\,) = 1$) yields red. $\chi^2$=1.0001 for the best fit of model 1 with 100\% $\langle100\rangle$-exchanges and red. $\chi^2$=0.994 for model 2 with 6\% $\langle100\rangle$ 44\% $\langle110\rangle$ and 50\% $\langle111\rangle$-exchanges.
It shall be noted, that when neglecting short-range order ($I_\text{SRO}(\vec{q}\,) = 1$) a model including 50\% $\langle111\rangle$-exchanges still agrees better with the experimental data than one exclusively including of $\langle100\rangle$ and $\langle110\rangle$-exchanges. 

%It is therefore , a significant number of $\langle111\rangle$-exchanges takes place at the temperature of 0.42$\times$T$_m$ in ordered Fe$_{0.54}$Al$_{0.46}$. 
%\clearpage
\subsection{Temperature dependence}
\label{sec:arrh}
Chemical diffusivities can be calculated from measurements at a certain $\vec{q}$-vector for different temperatures using the Einstein-Smoluchowski relation:
\begin{equation}
	\tilde{D}(T) = \frac{\left\langle R^2 \right\rangle}{6\tau_0} = \frac{\sum_n \bar{p}_n \left|r_n\right|^2}{6\tau_0}\, ,
\end{equation}
with $\bar{p}_n=p_n Z_n$ and $Z_n$ the coordination number. 

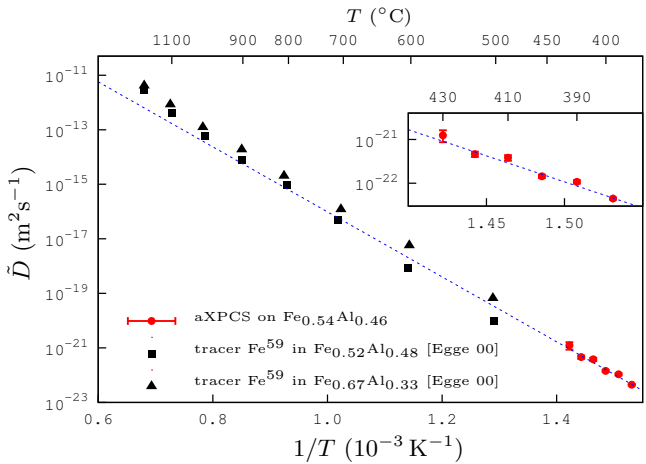
\begin{figure}[!t]
 	\centering
 	\begin{overpic}[width=0.5\textwidth]{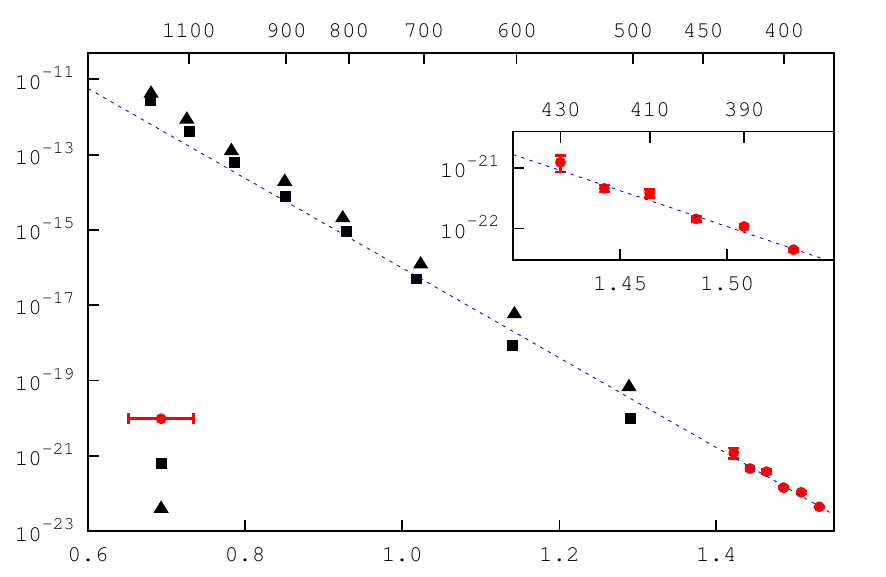}
		\put(97,-8){$1/T$ ($10^{-3}$\,K$^{-1}$)}
		\put(117,156){\footnotesize $T$ ($^\circ$C)}
		\put(-10,58){\rotatebox{90}{$\tilde{D}$ (m$^2$s$^{-1}$)}}
		\put(60,43){\tiny aXPCS on Fe$_{0.54}$Al$_{0.46}$}
		\put(60,31){\tiny tracer Fe$^{59}$ in Fe$_{0.52}$Al$_{0.48}$ \cite{Eggersmann2000}} 
		\put(60,19){\tiny tracer Fe$^{59}$ in Fe$_{0.67}$Al$_{0.33}$ \cite{Eggersmann2000}}	
	\end{overpic}
	\vspace*{0.3cm}	
 	\caption{Temperature dependence of chemical diffusivities $\tilde{D}$ calculated from aXPCS data using the model including $\langle111\rangle$-exchanges. For comparison, data from tracer measurements \cite{Eggersmann2000} in similar compositions is also shown. The blue line represents a fit to the aXPCS data according the Arrhenius law as shown in \eref{eq:arrh}. }
	\label{fig:arrh}
\end{figure}

Chemical diffusivities found at different temperatures are shown in \fref{fig:arrh} in comparison to self diffusion data found by tracer measurements \cite{Eggersmann2000}. 

The data measured in this study can be used to find the activation energy ($\tilde{E}_\text{a}$) of chemical diffusion by means of the Arrhenius equation:
\begin{equation}
	\tilde{D}(T) = \tilde{D}_0 \exp\left(\frac{\tilde{E}_\text{a}}{k_\text{B}T}\right),
	\label{eq:arrh}
\end{equation}  
where $k_\text{B}$ is the Boltzmann constant. The second model with jump probabilities \hbox{$ \bar{p}_{100} = 0.36$}, \hbox{$ \bar{p}_{110} = 0.12$} and $ \bar{p}_{111} = 0.52$ was applied, yielding a value of \hbox{$\tilde{E}_\text{a}=2.4(2)$\,eV}, which is in good agreement with the activation energy as a sum of formation and migration enthalpy of \hbox{$E_\text{a}=2.5(3)$\,eV} found by time-differential length change measurements for Fe$_{0.55}$Al$_{0.45}$ \cite{Schaefer1999}.   

%D_0=0.0000941755 + 0.00227468 - 0.0000904315
%Ea              = 2.36459          +/- 0.1876       (7.933%)
%DO              = 9.41754e-05      +/- 0.0003024    (321.1%)
%lnEa            = 2.36459          +/- 0.1876       (7.933%)
%lnDO            = -9.27035         +/- 3.225        (34.79%)

\section{Discussion}

It was shown that the dominating diffusion mechanism in B2 Fe$_{0.54}$Al$_{0.46}$ leads to effective $\langle111\rangle$-exchanges. 
A sequence of six correlated jumps of a vacancy, which is known in three different variations in a B2 ordered system, can lead to effective $\langle100\rangle$-jumps (there are two possibilities, a bent and a straight form) or to effective $\langle110\rangle$-jumps. If an antistructure atom is involved in the process this mechanism would lead to a corresponding exchange of two atoms of different kind along one of these two jump vectors. The six-jump cycle can therefore be ruled out as the dominating mechanism in this system at low homologous temperatures. In order to make an effective $\langle111\rangle$-jump possible, at least ten correlated jumps of one vacancy are required. As it is very unlikely that the energy landscape favors ten over six correlated jumps, mechanisms of this kind can be categorically ruled out as the dominating process of chemical diffusion in this system. 

The anti-structure bridge mechanism is a percolation dependent mechanism. A closed net of nearest-neighbor atoms of the majority type is necessary to allow for
nearest-neighbor exchanges of vacancies with majority atoms which does, however, not lead to chemical diffusion. 
The percolation threshold for B2 ordered systems was found to be 54.9(2)\,at.\% \cite{Divinski1997}. The composition under investigation here, therefore, lies just a little below this threshold. 
It is rather surprising that chemical diffusivities are similar to the self-diffusion of iron measured by a tracer method \cite{Eggersmann2000}. This can be seen as an important argument that the system is not fully percolated and the gaps between the percolated areas are bottlenecks for long-range diffusion.  

It can be assumed that the anti-structure bridge mechanism takes place in isolated regions, where the self diffusion constant would accordingly be much higher than the measured chemical diffusivity. 
In order to allow for material transport beyond the borders of the isolated, percolated clusters a mechanism bridging the gaps between these regions must take place. 

One possible explanation for such a mechanism would be the formation of a linear triple defect at the rim of a percolated cluster, where
two vacancies with an antistructure Fe-antiside atom in their middle along $\langle111\rangle$-direction exists. This defect could then diffuse by exchange its position with an Al-atom, leading to a transport of one Fe-atom from one percolated cluster to another. Extensive ab initio simulations are necessary to assess this hypothesis.     

\section{Conclusions}
We could show that aXPCS is capable of determining effective atomic exchanges and chemical diffusivities at temperatures unaccessible to other methods of diffusion studies with atomistic resolution. It was thus possible to study atomistic diffusion processes in a well ordered intermetallic for the first time. In the non-stoichiometric long-range ordered Fe$_{0.54}$Al$_{0.46}$ system at 0.42$\times$T$_m$ the dominating vacancy diffusion mechanism leads to effective $\langle111\rangle$-exchanges of Fe and Al atoms. The activation energy of chemical diffusion was found to be 2.4(2)\,eV.

%This led to the determination that the dominating diffusion path in non-stoichiometric long-range ordered Fe$_{0.54}$Al$_{0.46}$ as an effective $\langle111\rangle$-exchange, which allows to rule out the six-jump cycle as a diffusion controlling process in this system.  

\appendix

\section*{Acknowledgements}

This research was funded by the Austrian Science Fund (FWF) contracts P-22402 and P-28232. 

We thank S.~Puchegger from Faculty Center for Nano Structure Research University of Vienna for the EDX analysis. We also thank M. Sprung and F. Westermeier from beamline P10, PETRA III in Hamburg for a comprehensive support during our beamtimes.

Special thanks go to our collaborator Michael Leitner, now at Technische Universit{\"a}t M{\"u}nchen, Physik-Department E13, for many discussions and help in evaluating XPCS data.

\section*{Appendix}
The parameterization of the reciprocal space vector for elastic scattering and a sample oriented such that the (111) plain is perpendicular to the X-ray beam, \hbox{$[111] \parallel \vec{k}_\text{in}$}, with $k_0 = \left| \vec{k}_\text{in} \right|$ and $\hat{x},\hat{y},\hat{z}$ the Cartesian coordinates, is given by:
%\begin{figure*}
\newpage
\begin{strip}
{\small
	\begin{align}
 		\vec{q}\left(2\Theta, \varphi \right) =& 
 			2 k_0 \sin\left(\frac{2\Theta}{2} \right) \times   
			\Bigg(
 			\frac{\sin\left(2\Theta/2\right)}{\sqrt{3}} - \frac{\cos\left(2\Theta/2\right)}{6}\left(\sqrt{6}\cos(\varphi)+3\sqrt{2}\sin(\varphi) \right)
			\,\hat{x} \\ 
			&+ 
			\left(\frac{\sin\left(2\Theta/2\right)}{\sqrt{3}}+\frac{\cos\left(2\Theta/2\right)}{6}\left(-\sqrt{6}\cos\left(\varphi\right)+3\sqrt{2}\sin\left(\varphi\right)\right)\right)
			\,\hat{y} 
			+ 
			\frac{\sqrt{2}\cos\left(2\Theta/2\right)\cos\left(\varphi\right)+\sin\left(2\Theta/2\right)}{\sqrt{3}}
			\,\hat{z} \Bigg) \notag 
		\label{eq:q-param}
	\end{align}
}
\end{strip}
%\end{figure*}

%\newpage
%\section*{References}
%Literaturverzeichnis
%\bibliographystyle{hep}
\bibliographystyle{mod_wmaainf} %gerplain, gerunsrt, geralpha, gerapali
\bibliography{literature}
%\addcontentsline{toc}{chapter}{Literature}

\end{document}